\documentclass[fleqn,twoside]{article}
\usepackage{espcrc2}
\usepackage{graphicx}
\usepackage{epsfig}
\usepackage[figuresright]{rotating}
\def\taumg   {\tau \rightarrow \mu \gamma}

\def\mueg    {\mu \rightarrow e \gamma}
\def\taupair {\tau^+ \tau^-}
\def\mupair  {\mu^+ \mu^-}

\def\epair   {e^+ e^-}
\def\BABAR   {BABAR }
\def\mec     {m_{EC}}
\def\de      {\Delta E}

\newcommand{\AmS}{{\protect\the\textfont2
  A\kern-.1667em\lower.5ex\hbox{M}\kern-.125emS}}
\title{Search for the Lepton Number Violating Decay $\taumg$}

\author{Chris Brown\address[UVIC]{Department of Physics and Astronomy, 
        University of Victoria, \\ 
        P.O. Box 3055, STN CSC, Victoria BC V8W 3P6, Canada} for the \BABAR collaboration
}
\begin{document}
\begin{abstract}
Using data collected with the \BABAR detector between 1999 and 2001, we describe a preliminary search for the neutrinoless decay $\taumg$. This data sample includes data collected both on and off the $\Upsilon$(4S) resonance and corresponds to $56 \times 10^6$ produced tau pair events. The search has an efficiency of $5.2 \pm 0.1(MCstat) \pm 0.5(sys)\%$ and an expected background rate of $7.8 \pm 1.4$ events.  We select 13 events in the final sample. As there is no evidence for a signal in this data, we set a preliminary upper limit of B($\taumg$) $< 2.0 \times 10^{-6} @90\%$ CL.
\vspace{1pc}
\end{abstract}
\maketitle
\section{Introduction}
 
The decay $\taumg$ is an anticipated lepton-number violating process in supersymmetric models\cite{1,2,3}, left-right supersymmetric models\cite{4} and in supersymmetric string unified models\cite{5}. For some ranges of model parameters, decay rates as high several parts per million are expected for this decay\cite{3,5}, even in light of the current experimental limit on the related $\mueg$ decay\cite{6}. Assuming that neutrinos have mass, the standard model branching ratio is expected at the O($10^{-34}$) level. The current best published limit is B($\taumg$) $< 1.1 \times 10^{-6}$ from the CLEO experiment using 12.6 million $\tau$ pairs\cite{7}.

This analysis uses $56 \times 10^6$ $\taupair$ events from the 1999-2001 \BABAR data set to give a preliminary result of B($\tau \rightarrow \mu \gamma$).  The general signature of the $\taumg$ signal is the presence of an isolated $\mu$ and $\gamma$ which have an invariant mass consistent with that of the $\tau$(1.777 GeV) and with an energy in the centre-of-mass system (CMS) consistent with the beam energy in the CMS (5.29 GeV) and the rest of the particles in the event having properties that are consistent with being produced in a generic 1-prong $\tau$ decay\footnote{A `1-prong' decay contains a single charged particle amongst its decay products.}.  The analysis is performed using a blinded mass-energy region in the data which corresponds to a 3$\sigma$ error ellipse centred on the expected peak of the distribution. 

A cut-based approach is used, in which the background in the signal-box is estimated from extrapolations using side-band data and verified with the Monte Carlo simulation.  The determination of selection criteria in this analysis was based on Monte Carlo simulation of background and signal as well as on data in side-band regions further away from the signal-box. 

Background sources arising from non-$\taupair$ sources, the most problematic of which are radiative $\mupair$ pair events, are reduced to small levels. This leaves the generic 1-prong $\tau$ decays as the major source of background.   The dominant, and irreducible, background is from radiative muonic decays of the $\tau$ in which the two neutrinos have very little energy.
 
\section{Monte Carlo Samples}

The Monte Carlo simulation uses a complete description of the \BABAR detector response employing the GEANT4\cite{15} software.  The most important simulation samples are those involving $\epair \rightarrow \taupair$ decays.  These include 27.6 million generic $\taupair$ events, which use KORALB employing the TAUOLA decay package\cite{16}, and the $\taumg$ signal incorporated into TAUOLA.  Forty thousand signal events were generated which form the bulk of the signal Monte Carlo used for the analysis. A second signal sample with differing beam background contributions was produced in order to provide a cross check on the sensitivity of the selection to different background conditions.

Additionally, simulated $\mupair$ events, which uses the AfkQed generator\cite{17}, provide tools for $\mu$ particle identification and some guidance on $\epair \rightarrow \mupair$ event suppression. Other potential backgrounds include $\epair \rightarrow u\bar{u}; d\bar{d}; s\bar{s}, \epair \rightarrow c\bar{c}$ and $\epair \rightarrow b\bar{b}$ which were studied using the corresponding generic \BABAR simulated event samples.  As the selection is restricted to events in which both $\tau$'s decay via 1-prong modes, the $\epair \rightarrow b\bar{b}$ background is completely negligible as indicated by the simulated events. The $\epair \rightarrow c\bar{c}$ and light-quark background contributions are also found to be very small, with no events in the simulation surviving from these two sources after all cuts have been applied, and consequently have not been included further in the data-Monte Carlo comparisons. The two-photon processes are negligible backgrounds for events in which the beam energy is detected in a CMS hemisphere. 

\section{Mass and Energy Determination}

\begin{figure}[htb]
\includegraphics*[width=17pc,height=17pc]{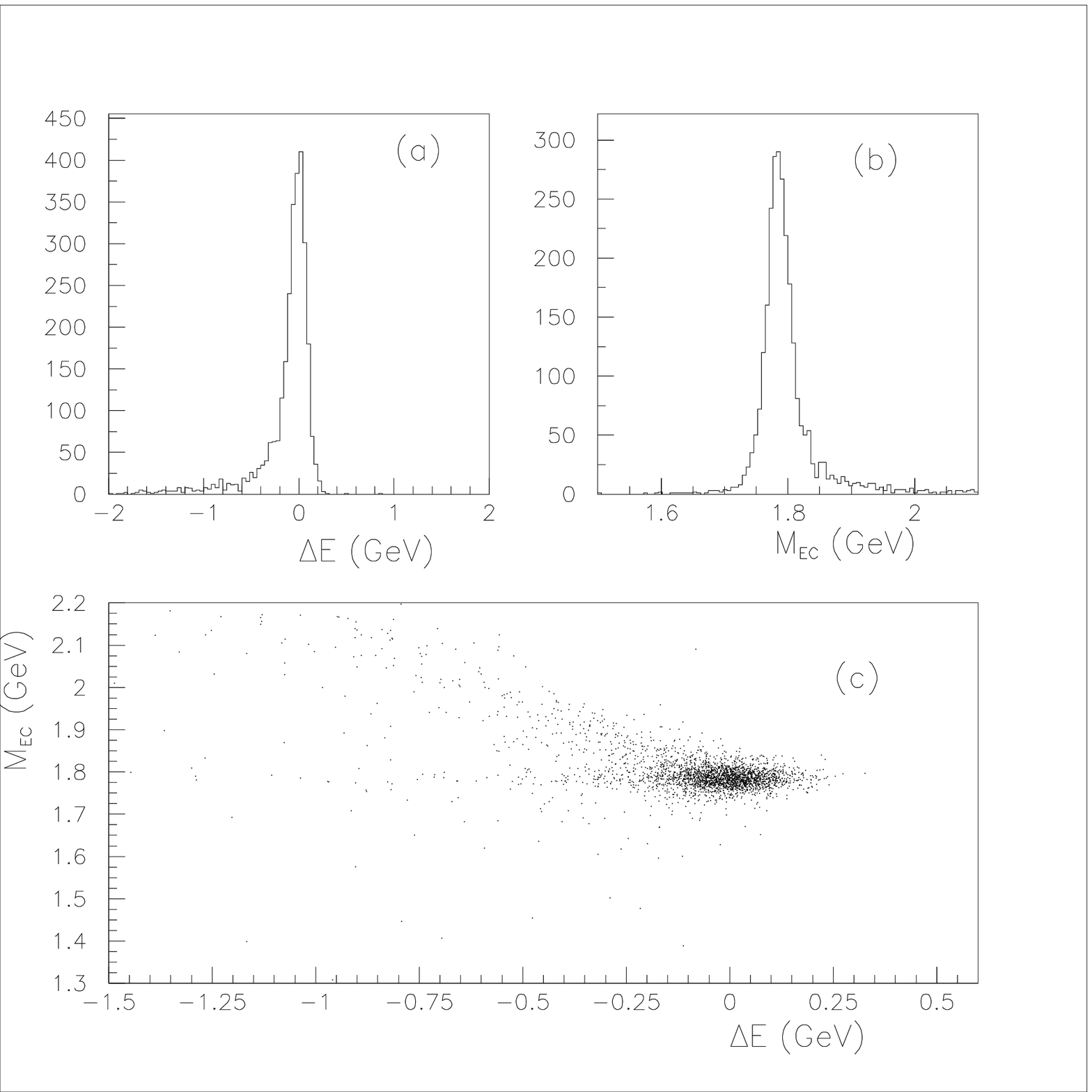}
\caption{Distributions of (a) $\de$ (b) $\mec$ and (c) $\mec$ vs $\de$ for the Monte Carlo Simulation of the $\taumg$ signal.}
\label{fig5}
\end{figure}
Because there is no missing energy in the $\taumg$ signal, the mass of the measured decay products is the $\tau$ mass and the energy is the full energy of the $\tau$. Ignoring the effects of initial-state radiation, the energy of the final state particles is equal to the full energy of the beam in the CMS. This situation lends itself very well to the use of a mass of the $\mu \gamma$ system calculated from the kinematic fit employing the beam energy constraint on the energy of the system as used by ARGUS, and denoted as $\mec$. The energy variable used is the difference between the measured $\mu \gamma$ energy and the beam energy, denoted by the symbol $\de$. The distributions of these variables are shown in Figure~\ref{fig5}(a)-(c). The resolutions of the core of these distributions, which represent those events with well reconstructed photons and tracks, are $\sigma_{\de}$ = 88 MeV and $\sigma_{\mec}$ = 19 MeV. 
 
However, as is evident by the diagonal band of events present in Figure~\ref{fig5}(c), when initial state radiation shifts the true energy of the $\tau$ from the beam energy, there a negative correlation between $\de$ and $\mec$ and a substantial loss of resolution.  The low-energy tail of the $\de$ distribution for well reconstructed $\mec$ , evident in Figure~\ref{fig5}(c), is populated by events with a photon reconstructed with the correct direction, but with significant energy loss. 
 
\begin{figure}[htb]
\includegraphics*[width=19pc,height=19pc]{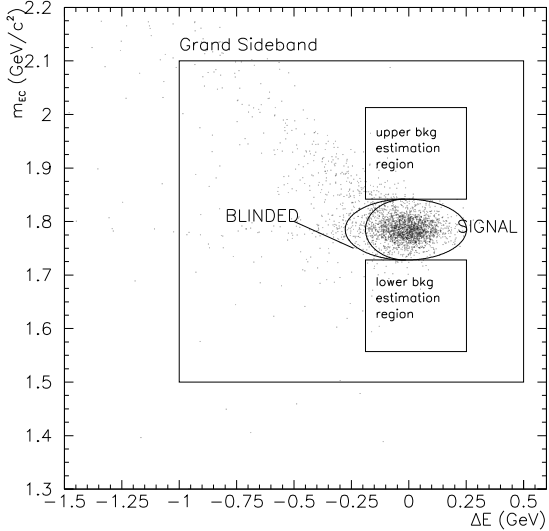}
\caption{Regions of the $\mec$ vs $\de$ plane shown schematically overlaying the distribution of the simulated signal.  The signal region is the assymetric ellipse within the elliptical blinded region.}
\label{fig7}
\end{figure}
Figure~\ref{fig7} shows the various regions in the $\mec$ vs $\de$ plane used for determining the selection criteria, measuring the background, and defining the signal and blinded regions. A `Grand Side-Band' is defined as the region within the $1.0 < \de < 0.5$ GeV and $1.5 < \mec < 2.1$ GeV bounds.  It is used for evaluating the reliability of the estimation of the signal efficiency and this is also shown on the figure.

The signal-box is an elliptical region centred on the peak of the two-dimensional distribution as determined by the Monte Carlo. For the positive side of $\de$, the ellipse has a 3$\sigma$ half-axis for both the $\de$ and $\mec$ axes whereas for the negative side of $\de$ the ellipse has a 2$\sigma$ half-axis in $\de$ and a 3$\sigma$ half-axis for $\mec$. The resolutions used in defining the signal-box are those of the core Gaussians obtained from the signal Monte Carlo.  This asymmetric shape provides an optimal signal-box given the resolution in $\de$ and the presence of increasing background in the negative $\de$ region that is not present for positive $\de$. 

Events that are ultimately selected and included in a study of the $\mec$ and $\de$ variables must first pass a number of selection criteria.  These requirements are designed primarily to reduce the number of Bhabha and $\mupair$ events.  The requirements of the standard \BABAR background filter which the signal events pass: 
\begin{small}
\begin{itemize}
\item 2 charged tracks
\item zero net charge
\item $|p_1| + |p_2| < 9.0$ GeV
\item $E_1 + E_2 < 5.0$ GeV
\item $E_1/|p_1| < 0.8$ OR $E_2/|p_2| < 0.8$
\item $E_{CM}-(|p_1|_{CM} + |p_2|_{CM}) > 0$
\item $p_{t(CM)}/(E_{CM}-(|p_1|_{CM} + |p_2|_{CM})) > 0.07$
\item charged track separation $> 90^\circ$ in the CMS 
\item at least one $\gamma$
\item fiducial region: $0.775 < \cos \theta_{track} < 0.940$
\end{itemize}
\end{small}
where $p_{1(2)}$ is the track momentum; $E_{1(2)}$ is the calorimeter energy associated with the track; $p_{t(CM)}$ is the total transverse momentum of charged tracks in the CMS; and subscript CM indicates when the quantity is in the CMS. 

These requirements have an efficiency of $37.3 \pm 0.2\%$.  These efficiency losses arise largely from the fiducial acceptance of the detector, but also from the branching ratio to 1-prong modes. At this stage of the analysis, $22 \times 10^6$ events are selected in the data.  Monte Carlo studies indicate that 83\% of the signal events pass these requirements for events which would have otherwise been selected.

Subsequent selections employ requirements to remove the non-$\tau$ background sources and the $\epair \rightarrow \mupair \gamma$  and $\epair \rightarrow \epair$ events in particular.  Cuts include a specific requirement on the observed CMS energy, $E^{tot}_{vis}/2E^{CMS}_{Beam} < 0.95$,  to exclude the aforementioned background sources in general and a $\mu$ veto for reduction of di-muon backgrounds in particular.  The event is tagged via an electron or $h \geq 1 \pi^0$ on the opposing side.  The missing mass on the tag side is restricted to reduce backgrounds from lost or difficlt to reconstruct tracks.  To be consistent with a decay through an intermediate $\tau$, the tag track is required to have less than 80\% of the CMS beam energy.  In general, this will not be the case for the non-$\tau$ backgrounds.

On the signal side, we impose the requirement of an identified $\mu$ and an associated $\gamma$ with energy in excess of 400 MeV.  These cuts are applied in addition to the obvious $\mec$ and $\de$ cuts for the final selection.   

The cut progression shows that once the non-$\tau$ background is removed by requiring an electron-vs-$\mu \gamma$ or $h \ge 1 \pi^0$-vs-$\mu \gamma$ events, the $\tau$ Monte Carlo simulation tracks the data reasonably well.

\begin{figure}[htb]
\includegraphics*[width=7cm,height=7cm]{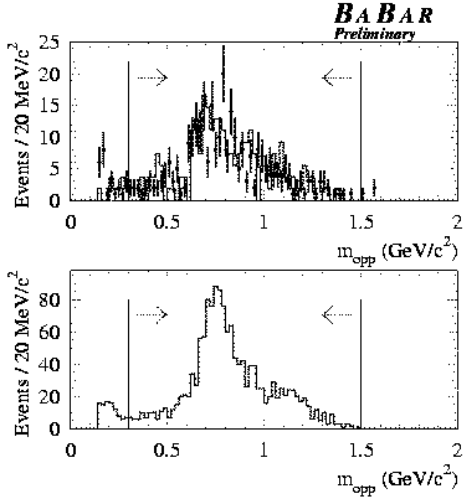}
\caption{Distributions of $m_{opp}$ after all other requirements, apart from the final $\mec$-$\de$, have been applied.  The data and $\tau$ Monte Carlo are in the top plot and the $\taumg$ signal Monte Carlo in the bottom plot.  The parts of the distributions accepted in the selection are indicated.}
\label{fig14}
\end{figure}

\begin{figure}[htb]
\includegraphics*[width=7cm,height=7cm]{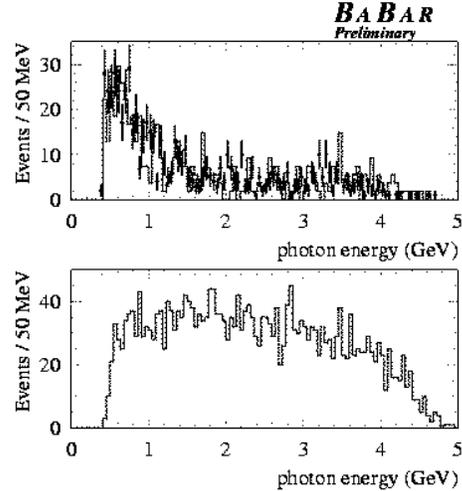}
\caption{Distributions of $E_\gamma$ after all other requirements, apart from the final $\mec$-$\de$, have been applied.  The data and $\tau$ Monte Carlo are in the top plot and the $\taumg$ signal Monte Carlo in the bottom plot.  }
\label{fig12} 
\end{figure}

We present in Figures~\ref{fig14} and~\ref{fig12} a sample of the distributions of variables employed in the analysis after all other requirements, apart from that using the one shown, have been applied. The data distributions are represented by points and the $\tau$ Monte Carlo simulation by histograms in these figures. The normalizations of the Monte Carlo distributions is fixed by the luminosity.

These distributions are well described by the $\taupair$ simulation in both shape and overall normalization. There are 604 events in the data in these distributions and the ratio of data to the Monte Carlo expectation is $1.022 \pm 0.069(stat) \pm 0.025(norm)$. The first error is statistical and the second is the systematic error associated with normalization of the Monte Carlo events. This normalization uncertainty is a component of the systematic error associated with the efficiency of selecting the signal. This includes uncertainties in integrated luminosity, cross-section, radiation treatment in the generator and branching ratio uncertainties. 

Because the signal has a relatively stiff momentum spectrum, muon particle identification ($\mu$-PID) efficiency using a sample of radiative $\mu$ pair events ($\epair \rightarrow \mupair\gamma$) from the same data set used for the rest of the analysis was exploited to study the $\mu$-PID efficiency. Contamination from sources other than the $\mu$ pair, such as $\taupair$ events, are negligible in this sample. 

The ratio of the efficiencies as a function of LAB momentum of the data to that of the $\epair \rightarrow \mupair(\gamma)$ simulation was studied and a correction based upon the control sample was extracted. Similarly obtained is the correction for the $\taupair$ events in the Monte Carlo sample and the signal Monte Carlo.  This method for applying the correction is particularly appropriate here as the polar-angle, azimuthal and momentum distributions and their correlations are very similar between the $\mu$ pair control sample and the signal and background. From these studies, an additional correction factor of 0.80 to 0.85 is applied to the efficiency as predicted by the Monte Carlo. 

\section{Backgrounds}

There are two broad categories of background that require suppression: the background arising from non-$\taupair$ sources, the most significant being $\epair \rightarrow \mupair (\gamma)$, and those coming from standard $\tau$ decays. In the latter category, the $\tau \rightarrow \mu \nu \nu \gamma$ form an irreducible background. 

The $\tau$ backgrounds are supressed by requiring the signal $\gamma$ to be energetic, $E_\gamma > 40 MeV$, and by imposing tight requirements on the $\mu$-PID of the track.

The non-$\tau$ background is reduced by requiring there to be a non-$\mu$ tag on the non-signal hemi-sphere and by removing events with measured momentum and calorimeter energy characteristic of events and hemispheres with little or no undetected energy. These events are particularly problematic as they naturally possess a value of $\de$ that overlaps the signal. 
 
After all the selection criteria have been applied, before applying the signal-box cut, there remain background sources in the Grand Side-Band region that, from the Monte Carlo simulation, consist of  $\tau \rightarrow \mu \nu \bar{\nu}$ (85.9\%), $\tau \rightarrow \pi \nu$ and $\tau \rightarrow K \nu$ (10.6\%) and $\tau \rightarrow \rho \nu$ (3.5\%). The non-$\tau$ background is expected to be very small in the final sample. 

The background is estimated by fitting the side-bands in $\mec$ on a sample of data selected to have $-2\sigma_{\de} < \de < 3\sigma_{\de}$. As seen in Figure~\ref{fig83}, the distribution of the background is reasonably uniform in the region of the final selection, which enables one to use a simple linear interpolation to estimate the density of background one expects in the signal-box region.  A geometrical correction is applied to estimate the background contained in the elliptical signal-box. 

A low statistics check of the method is made by applying it to the $\taupair$ Monte Carlo sample. In this case, the luminosity scaled number of events predicted from the side-bands is $9.1 \pm 2.1$ events, in good agreement with the $5.5 \pm 3.2$ events observed. This estimate of $5.5 \pm 3.2$ is the only background estimate based purely on counting Monte Carlo events.
 
The data side-band measurements yield estimates of $7.8 \pm 1.4$ events.

\section{Systematics}

Using the $\taupair$ Monte Carlo, the above selections yield an absolute efficiency for the signal of $6.7 \pm 0.1(stat)\%$. After the final selection of requiring the events fall within the $\mec$ - $\de$ signal-box, the efficiency is $5.2 \pm 0.1(MCstat) \pm 0.5(sys)\%$. This efficency has a number of systematic uncertainties associated with it including those arising from: 
\begin{small}
\begin{enumerate}
\item the trigger efficiency 
\item the tracking reconstruction efficiency 
\item the neutral cluster reconstruction efficiency 
\item the background filter and skim selection efficiency 
\item electron opposite hemisphere requirements (electron tag) 
\item $m_{opp}$ requirements (`$\rho$' tag) 
\item $\mu$-PID requirements 
\item the photon energy scale and resolution 
\item the photon direction reconstruction, scale and resolution 
\item the track momentum scale and resolution 
\item the track momentum direction scale and resolution 
\item the beam energy scale 
\item the beam energy spread 
\end{enumerate}
\end{small}
Evaluation of the efficiency done using the events in the Grand Side-Band, as these have characteristics which are very similar to those of the signal.  The good agreement between the data and Monte Carlo normalization for the background sources, evident in Table 1, and the shapes of, for example the $\mu$ momentum, indicates that the efficiencies are well understood. The effects of systematic items (1)-(7) are incorporated into the ratio of observed to expected events: $1.022 \pm 0.069(stat) \pm 0.025(norm)$.  This yields an estimate of 7.3\% on the systematic error associated with items (1)-(7). 

A global estimate of tracking and calorimetry errors is provided by shifting the signal-box position and by modifying the size of the box according to the uncertainties in the resolution. Half of the observed changes are used to estimate the systematic errors.  The systematic errors assessed in this manner are presented in Table~\ref{table2}. 

Monte Carlo simulations revealed that the selection efficiency is insensitive to the uncertainties in the beam energy spread and the associated systematic error is negligible. 

\begin{table}
\begin{tabular}{|l|c|}
\cline{1-2}
 Systematic Influence &  Relative Error\\
 on the Signal Efficiency & (\%)\\
\cline{1-2}
 Effects (1)-(7) &  $\pm 7.3$\\
\cline{1-2}
Track and Ecal Resolution: &  \\
\cline{1-2}
$\de$ scale & $\pm 0.8$ \\
$\de$ resolution & $\pm 3.4$ \\
$\mec$ scale & $\pm 0.3$ \\
$\mec$ resolution & $\pm 0.6$ \\
Ecal scale & $\pm 3.3$ \\
Momentum scale & negligible \\
beam energy spread & $\pm 0.3$ \\
\cline{1-2} 
Total & $\pm 8.8$ \\
\cline{1-2}
\end{tabular}
\caption{ Systematic Errors associated with the signal efficiency.}
\label{table2}
\end{table}

\section{Limit}

From the observed number of events a 90\% upper limit is set on B($\taumg$) with the systematic errors included as suggested in \cite{9} using the technique of \cite{rb}. 

A demonstration with the two sidebands can be made: the lower sideband had a predicted background rate of 6.1 $\pm$ 2.2 events, whereas 6 are observed. 

\begin{figure}[htb]
\includegraphics*[width=7cm,height=7cm]{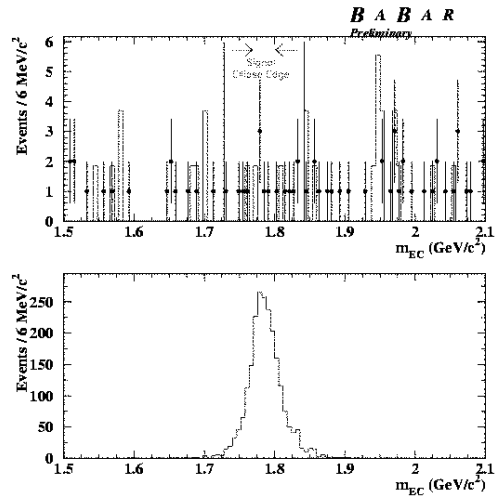}
\caption{Distribution of $\mec$ for the data and $\taupair$ Monte Carlo simulation for the $\taumg$ simulated signal for those events within the $-2\sigma_{\de} < \de < 3\sigma_{\de}$ region.  This is after all cuts but that on the signal-box.  This plot is made after unblinding.}
\label{fig81} 
\end{figure}
\begin{figure}[htb]
\includegraphics*[width=7cm,height=7cm]{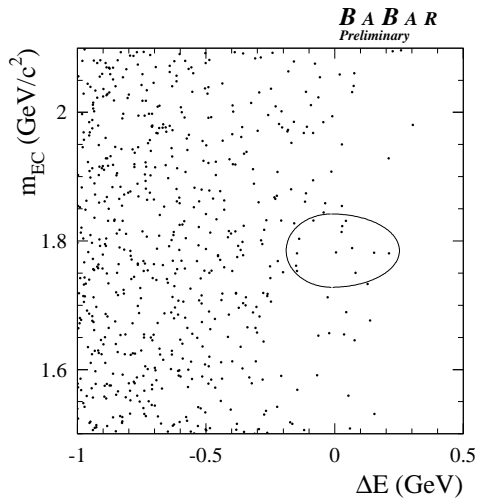}
\caption{Distribution of $\mec$ vs $\de$ for data.  The signal-box is indicated as the assymetric ellipse and shows the 13 observed events.}
\label{fig83} 
\end{figure}

When the signal-box was unblinded, we observe 13 events, shown in Figure~\ref{fig81} for the $\mec$ projection. Of these, 9 had electrons identifed in the opposite hemisphere; 5 were $h\geq\pi^o$-like and one was identified as both an electron and a $h\geq\pi^o$. These tagging ratios are consistent with the expectations of $\tau$ decays. The distributions of the $\mec$ and $\de$ are shown in Figure~\ref{fig83}.   

The background was estimated to be $7.8 \pm 1.4$. This yields a limit of 11.5 event upper limit on the number of signal events @90\% CL when the systematic errors are included. The 11.5 event upper limit translates into a limit: B($\taumg) < 2.0 \times 10^{-6} @90\%$CL. The probability of a background of $7.8 \pm 1.4$ events up to 13 observed events in the absence of a signal is 7.6\% if one includes the systematic errors. 

\section{Conclusion}

The 1999-2001 \BABAR data has been studied in a search for the forbidden decay $\taumg$. These studies reveal that a search with an efficiency of $5.2 \pm 0.1(MCstat) \pm 0.5(sys)\%$ and an expected background rate of $7.8 \pm 1.4$ results in 13 events being observed. This leads to a preliminary limit of B$(\taumg) < 2.0 \pm 10^{-6}@90\%$CL.

\end{document}